\documentclass[preprint,prl]{revtex4}
\usepackage{amsfonts}

\input{tcilatex}

\begin{document}

\section{Difficulty in the Fermi-Liquid-Based Theory for the In-Plane
Magnetic Anisotropy in Untwinned High-$T_{c}$ Superconductor}

Recently, Eremin and Manske \cite{Eremin-Manske} presented a one-band
Fermi-liquid theory for the in-plane magnetic anisotropy in untwinned high-$%
T_{c}$ superconductor YBa$_{2}$Cu$_{3}$O$_{6.85}$ (YBCO). They claimed that
they found good agreement with inelastic neutron scattering (INS) spectra 
\cite{hinkov}. In this comment, we point out their conclusion on this
important problem maybe questionable due to an error in logic about the
orthorhombicity $\delta _{0}$ characterizing the lattice structure of YBCO.
In Ref. \cite{Eremin-Manske}, a single band at $\delta _{0}>0$ is proved to
be in accordance with the angle resolved photoemission spectroscopy (ARPES)
on untwinned YBCO. But in its Erratum \cite{Erratum}, they admit that $%
\delta _{0}=-0.03$ was used to fit the INS data. Hence publications \cite%
{Eremin-Manske,Erratum} contain errors that we believe invalidate their
approach.

Let's make clear how $\delta _{0}>0$ was proved in the subject Letter.
Briefly a positive $\delta _{0}$ leads to in-plane anisotropic hopping
integrals $t_{a}>t_{b}$ from $t_{a}/t_{b}=(1+\delta _{0})/(1-\delta _{0})$.
This is consistent with $t_{a}/t_{b}\approx (b/a)_{{}}^{4}$ by \textit{%
ab-initio} calculation \cite{lichti} noting the lattice constants $a<b$ in
YBCO, which was cited in Ref. \cite{Eremin-Manske} to support $\delta
_{0}=0.03$. Furthermore, the ARPES data on untwinned YBCO \cite{schabel}
were used in Ref. \cite{Eremin-Manske} to prove a positive $\delta _{0}$.
Eremin and Manske have correctly argued that a suppression of the ARPES
intensity observed around $(\pm \pi ,0)$ due to absence of the van Hove
singularity is consistent with Fermi surfaces closing around $(\pm \pi ,0)$
as shown in Fig. 1 (with $\delta _{0}>0$) of Ref. \cite{Eremin-Manske}.
Positive $\delta _{0}$ based on ARPES data was adopted in Ref. \cite{PRB
2004}, which was cited in \cite{Eremin-Manske,Erratum}.

Unfortunately, Eremin and Manske need $\delta _{0}=-0.03$ in their Erratum 
\cite{Erratum}. They made significant modification of Fig. 1 through
rotating the Fermi surfaces by 90%
${{}^o}$%
. It is immediately obvious that the negative $\delta _{0}$ adopted by \cite%
{Erratum} breaks the right physical arguments in \cite{Eremin-Manske}.
Especially, the new Fermi surfaces closing around $(0,\pm \pi )$ \cite%
{Erratum} contradict the ARPES experiment cited in Ref. \cite{Eremin-Manske}.

Another important physical error in the subject Letter about the
superconducting gap, $\Delta _{k}=\delta _{0}\Delta _{s}+(1-\delta
_{0})\Delta _{d}$ with $\Delta _{d}=$ $\Delta _{0}(\cos k_{x}-\cos k_{y})/2$
and $\Delta _{s}=\Delta _{0}(\cos k_{x}+\cos k_{y})$, should also be pointed
out. The ARPES experiment on untwinned YBCO has observed obvious difference
for the maximum gaps at $(0,\pm \pi )$ and $(\pm \pi ,0)$ as cited in Ref. 
\cite{Eremin-Manske}. On the contrary, the values of $\Delta _{k}$ at these
two corners are $\pm \Delta _{0}(1-\delta _{0})$, respectively. In other
word, the gaps are actually equal in magnitudes. Thus the superconducting
gap assumed in Ref. \cite{Eremin-Manske} is lack of\ physical ground.

The orientation of the rectangular INS pattern \cite{hinkov} in untwinned
high-$T_{c}$ superconductor YBCO is of fundamental importance. To some
extent, success in explanation of this experiment may distinguish different
theoretical scenarios of high-$T_{c}$ cuprate superconductors as claimed in
the subject Letter. But there is the contradiction between explanations of
ARPES and INS data in untwinned high-$T_{c}$ superconductor YBCO in Ref. 
\cite{Eremin-Manske,Erratum}, which invalidates their one-band Fermi-liquid
theory. This is why we believe it is valuable to point out the sign error
about $\delta _{0}$\ in Ref. \cite{Eremin-Manske,Erratum}. This difficulty
makes the conventional solution in the subject Letter questionable \cite{3
PRB}.

\bigskip This work was supported by NSFC (10574063, 10474032).\bigskip

\bigskip Hou-Jun Zhao and Jun Li

\bigskip National Key Laboratory of Solid State Microstructures and
Department of Physics, Nanjing University, Nanjing 210093, PR China.

\end{document}